# Superconductivity in $LiFeO_2Fe_2Se_2$ with anti-PbO-type Spacer Layers

X. F. Lu[1], N. Z. Wang[1], G. H. Zhang[2,3], X. G. Luo[1], Z. M. Ma[3], B. Lei[1], F. Q. Huang[2,3], X. H. Chen[1]

1. Hefei National Laboratory for Physical Sciences at Microscale and Department of Physics, University of Science and Technology of China, Hefei, Anhui 230026, China

2. CAS Key Laboratory of Materials for Energy Conversion, Shanghai Institute of Ceramics, Chinese Academy of Sciences, Shanghai 200050, China

3. Beijing National Laboratory for Molecular Sciences and State Key Laboratory of Rare Earth Materials Chemistry and Applications, College of Chemistry and Molecular Engineering, Peking University, Beijing 100871, China

Superconductivity has been found in the iron-arsenides with different structures by alternative stacking of the conducting $Fe_2As_2$ layers and spacer layers including alkali- or alkali-earth metal ions and PbO- or perovskite-type oxides blocks[1-7]. So far, no spacer layer could be intercalated in between the neutral $Fe_2Se_2$ layers similar in structure to $Fe_2As_2$ layers except for alkali-metal ions in $A_xFe_{2-y}Se_2$[8]. The superconducting phase in $A_xFe_{2-y}Se_2$ with transition temperature of 32 K is always inter-grown with the insulating phase which exhibits an antiferromagnetic order with extremely high Neel temperature of ~560 K and the order of Fe vacancies[9,10]. Such inhomogeneity makes the investigation on the underlying physics in $A_xFe_{2-y}Se_2$ complicated. Here we report the synthesis of a novel superconductor $LiFeO_2Fe_2Se_2$ by hydrothermal method, exhibiting superconductivity at ~43 K, in which anti-PbO-type spacer layers of $LiFeO_2$ have been successfully intercalated between anti-PbO-type $Fe_2Se_2$ layers. This finding demonstrates that superconductivity can be realized in the iron selenide with a novel spacer layer of anti-PbO-type, which is not found in the iron arsenides, and expands the category of the iron-based superconductors. Such a new synthetic method paves a new way to search for possible novel superconductors with different spacer layers, which is helpful to further study the underlying physics in iron-based high-$T_c$ superconductors.

β-FeSe has the same basic structure as the FeAs layer in iron arsenide superconductors, and shows superconductivity at ~8 K[11]. However, there exists a remarkable difference between them, that is, the $Fe^{2+}Se^{2-}$ layer is neutral, while $Fe^{2+}As^{3-}$ layer has negative charge. It is the electric neutrality that makes the physical properties and structural category of FeSe-derived compounds different from that of the FeAs-derived compounds. In iron arsenide superconductors, the structure is formed by the alternative stacking of conducting FeAs layers and various spacer layers including alkali-metal ions (*A*FeAs; *A*=Li, Na)[5,12], alkali-earth metal ions (*Ae*Fe$_2$As$_2$; *Ae*=Ba, Sr, Ca, K, Cs)[3,4], PbO-type oxides (*Ln*OFeAs; *Ln*=La, Ce, Pr, Nd, Sm, etc.)[1,2,13], and perovsite-type oxides ((Ca$_3$*M*$_2$O$_{5-y}$)(Fe$_2$*Pn*$_2$); *M*=Sc, V, Al, etc., *Pn*=As, P; (Ca$_4$*M*$_2$O$_{6-y}$)(*Fe*$_2$Pn$_2$); *M* =Sc, V, Ti, Al, etc., *Pn*=As, P)[6,7]. The superconductivity with highest $T_c$ in FeAs-derived compound has been realized in the *Ln*OFeAs (1111 phase) with the spacer layers of PbO-type oxides[14]. In iron selenides, the alkali metal ions ($A^+$) as spacer layer can be only inserted into between FeSe layers to form $A_x$Fe$_{2-y}$Se$_2$ so far[8,15,16]. $A_x$Fe$_{2-y}$Se$_2$ single crystal contains two distinct phases: a superconducting phase with $T_c$ of 32 K and no Fe vacancies[9], and an insulating antiferromagnetically ordered phase of K$_2$Fe$_4$Se$_5$ with Neel transition temperature of ~560 K and √5×√5 order of Fe vacancies[10]. However, the superconducting phase can be observed only in single crystals and is always inter-grown with the insulating phase, and there is no evidence of this phase in polycrystalline samples. It seems that the insulating phase $A_2Fe_4Se_5$ closely correlates with the superconductivity in $A_xFe_{2-y}Se_2$, and even is believed to be the parent compound[17]. In order to acquire a better understanding of the underlying physics in iron-based high-$T_c$ superconductors, it is of significance to find novel spacer layer to expand the category of the iron-based superconductors. Here, we report a novel iron selenide superconductor $LiFeO_2Fe_2Se_2$ with neutral anti-PbO-type spacer layers synthesized by a new synthetic method. As shown in Fig. 1a, the structure of $LiFeO_2Fe_2Se_2$ is formed by alternatively stacking of anti-PbO-type $Fe_2Se_2$ and $LiFeO_2$ layers. It is emphasized that $LiFeO_2$ is anti-PbO-type spacer layer and has cations-disordered cubic rock-salt structure (so called α-$LiFeO_2$)[18], and different in structure from the PbO-type $LnO$ layers in $LnOFeAs$. In $LiFeO_2$ layers, Li and Fe randomly occupy half by half at the same site, which makes the $LiFeO_2$ layers electrically neutral. Generally speaking, the charge neutral nature of $Fe_2Se_2$ and $LiFeO_2$ layers retrains charge transfer between them, and the $LiFeO_2$ layers should serve only as spacer blocks.

Fig. 1b shows the X-ray diffraction (XRD) pattern of as-synthesized $LiFeO_2Fe_2Se_2$ taken at room temperature and the refinement fitting of Rietveld analysis[19] with a structural model shown in

Fig. 1a, which adopts alternatively stacking of anti-PbO-type $Fe_2Se_2$ and anti-PbO-type $LiFeO_2$ layers. All Bragg peaks of the product were well indexed to a tetragonal structure with space group *P*4/*nmm* (No. 129). Table 1 lists the final crystallographic parameters from the refinement, and Fig. 1b shows Rietveld refinement patterns. The obtained lattice constants of the tetragonal $LiFeO_2Fe_2Se_2$ are $a$ = 3.7926(1) Å and $c$ = 9.2845(1) Å, which has a much smaller *a*-axis lattice parameter but larger *c*-axis lattice parameter compared with *Ln*OFeAs (for instance, $a$ = 4.0337 Å, $c$ = 8.7411 Å for LaOFeAs[20]), respectively. The final *R* factors is fairly small ($R_{wp}$=0.0937), which gives evidence for the validity of our present structural model. The unique bond distance of Fe-Se is 2.4347(1) Å and the unique bond distance of Li(Fe)-O is 2.0246(1) Å. The rather large distance of 3.6028(1) Å between Se site and O site indicates no chemical bonds (van der Waals gap) between neighboring $Fe_2Se_2$ and $LiFeO_2$ layers. The two unique angles of Se-Fe2-Se are 102.31° and 113.16°, respectively.

The magnetic susceptibility ($\chi$) as a function of temperature for $LiFeO_2Fe_2Se_2$ is plotted in Fig. 2. The measurements were performed under an external field $H$ = 10 Oe. As shown in Fig.2a, the as-synthesized sample shows a round diamagnetic transition around 40 K (see the inset), which is higher than $T_c$ (~37 K) observed in FeSe under pressure[21] and 32 K in $A_xFe_{2-y}Se_2$[8,15,16], respectively. Fig.2a shows a considerable shielding fraction of 44% at 5 K in the zero-field cooling process. Such a large diamagnetism indicates that the as-synthesized $LiFeO_2Fe_2Se_2$ sample undergoes bulk superconductivity with superconducting transition around 40 K. In low-temperature solution synthetic route, the small sizes of the crystalline grains in product can lead to a low superconducting shielding fraction. To further enhance the sample quality, we adopted high-pressure annealing technique to get rid of the residual solution and improve occupation random of atoms. Fig. 2b exhibits the magnetic susceptibility of the $LiFeO_2Fe_2Se_2$ sample annealed under high pressure of 5 GPa and at 150 $^o$C for 5 hrs. The sharp drop of χ at ~43 K suggests that the high-pressure annealing procedure enhanced the superconducting transition temperature ($T_c$) effectively. Indeed, our data indicates that the annealing procedure not only enhanced $T_c$ but improved the diamagnetism. The χ of the annealed sample at 5 K taken in the field cooling process reaches 36% of theoretical value of perfect diamagnetism, which is more than 10 times enhanced compared to as-synthesized sample. The χ of the annealed sample at 5 K taken in the zero-field cooling process shows that the superconducting shielding fraction reaches 53%. The inset shows the M-H loop taken at 5 K for the annealed sample. It indicates that $LiFeO_2Fe_2Se_2$ is a type-II superconductor. A linear-*H* dependence of diamagnetic susceptibility with negative slope can be recognized in low-field range below about 300 Oe, which further confirms that the steep decrease of $\chi$ originates from the superconducting transition, and a lower critical field $H_{c1} \approx 300$ Oe at 5 K can be inferred.

Fig. 3a presents the magnetization of annealed $LiFeO_2Fe_2Se_2$ under various pressures ranging from 0 – 1 GPa and $H = 10$ Oe in zero-field cooling process. $T_c$ monotonically decreases with increasing the external pressure, indicating the suppression effect of pressure on superconductivity in $LiFeO_2Fe_2Se_2$. Fig. 3b summarizes the pressure dependence of $T_c$. As the external pressure increases from 0 to 1 GPa, $T_c$ decreases by almost 5 K with a negative slope $d(T_c/T_c(0))/dP \sim -0.12$ $GPa^{-1}$. This value is larger than that observed in $K_xFe_{2-y}Se_2$ single crystals[22]. This difference could be related to the much larger height of anion from Fe site ($h$) in $LiFeO_2Fe_2Se_2$ (1.527 Å, inferred from the refined parameters in Table 1) than that in $K_xFe_{2-y}Se_2$ (1.459 Å, calculated from the crystallographic data in Ref. 7) because $T_c$ is closely related to the chalcogen height[23], which is easily suppressed by pressure[24].

It has been found that $T_c$ in the FeAs-derived superconductors is closely correlated to the anion height ( $h$ ) from Fe layer within one $Fe_2Pn_2$ ($Pn$ = As, P) layer[23]. As shown in Fig.1 of Ref. 23, $T_c$ displays a maximum at $h_0 \approx 1.38$ Å and decreases rapidly with deviation of $h$ from $h_0$. However, this rule is not suitable for the FeSe derived superconductors. $T_c$ increases from 8 K in FeSe to 43 K in $LiFeO_2Fe_2Se_2$ with increasing $h$ from 1.45 Å[25] to 1.527 Å, respectively. In addition, the anion height of 1.527 Å in LiFeO$_2$Fe$_2$Se$_2$ is nearly the same as 1.519 Å in Li$_{0.6}$(NH$_2$)$_{0.2}$(NH$_3$)$_{0.8}$Fe$_2$Se$_2$[26], and they share the same $T_c$ of 43 K. It seems to indicate that the $T_c$ is also related to the anion height in iron selenide superconductors. In Fig.4, we plot $T_c$ as a function of chalcogen (Se) height for all the FeSe-derived superconductors. The phase diagram shows two trends. One is that $T_c$ monotonically increases with deceasing Se height less than 1.45 Å; while the second one is that $T_c$ quickly increases with increasing the height larger than 1.45 Å. Following the second trend, it is easy to understand that $T_c$ of LiFeO$_2$Fe$_2$Se$_2$ monotonically decreases under pressure as shown in Fig.3, as well as in K$_x$Fe$_{2-y}$Se$_2$[22,24], because the Se height decreases with pressure, being in contrast to the case of FeSe with the height less than 1.45 Å. It indicates different physical behavior for the height larger and less than 1.45 Å. It should be addressed that all the FeSe-derived superconductors with the Se height larger than 1.45 Å is heavily electron doped. In this sense, it is possible to achieve the higher Tc in LiFeO$_2$Fe$_2$Se$_2$ by changing the ratio of Li/Fe, and consequently carrier concentration to adjust the Se height and to control $T_c$. In addition, it is also reported that extension of the lattice in *ab* plane leads to an increase of $T_c$[27]. In one word, it is possible to achieve higher $T_c$ by adjusting the anion height or lattice constant of *ab* plane in novel FeSe-derived superconductors with different spacer layers.

As reported in Ref. 28, α-LiFeO$_2$ shows antiferromagnetism with a large effective moment of 4.53(3) $\mu_B$ and Weiss constant of -186(3) K, and the magnetic transition temperature of α-LiFeO$_2$ is governed by the degree of cation ordering. Therefore, it is quite significant to study magnetic correlation between Fe$_2$Se$_2$ and LiFeO$_2$ layers for understanding the mechanism of high-Tc superconductivity because it is widely believed that the cooper pairing originates from the spin fluctuation in the iron-based high-$T_c$ superconductors[29], as the same as that in high-$T_c$ cuprate superconductors[30].

In the present work, we fabricate a novel FeSe-based layered material LiFeO$_2$Fe$_2$Se$_2$ through a new synthetic route (hydrothermal reaction method), in which an anti-PbO-type LiFeO$_2$ block was intercalated between Fe$_2$Se$_2$ layers. We discovered that the superconductivity therein can reach a critical temperature as high as 43 K, being much higher than 8 K of FeSe. This work expands the category of the iron-based superconductors and opens a new window to achieve iron-based superconductors for higher $T_c$.

**Method:** Polycrystalline samples of (Li$_{0.5}$Fe$_{0.5}$)OFeSe were prepared by hydrothermal reaction method. 0.012-0.02 mol selenourea (Alfa Aesar, 99.97% purity), 0.0075 mol Fe powder (Aladdin Industrial, A.R. purity), and 12 g LiOH·H$_2$O (Sinopharm Chemical Reagent, A.R. purity) were put into a Teflon-lined autoclave (50 mL) and mixed together with 10 ml deionized water. The Teflon-lined steel autoclave was following tightly sealed and heated at 160 ºC for 3-10 days. The iron powder was hydrolyzed, and then FeO(OH), as the hydrolyzate, was partly reduced due to the weakly reducibility of selenourea. FeSe was obtained from the combination of reduced $Fe^{2+}$ ion and the selenourea, and meanwhile $Li^+$ ion partially substitutes $Fe^{3+}$ ion in FeO(OH). Finally the reaction between FeSe and $FeLiO_2$ would produce $FeLiO_2Fe_2Se_2$. This is possible reaction mechanism. The polycrystalline samples acquired from the reaction system with shiny lamellar were washed with deionized water repeatedly, and dried in room temperature. Atomic ratio of Li: Fe: Se is determined to be 1: 3: 2 by inductively coupled plasma-atomic emission spectroscopy (ICP-AES), with the instrument error of 10%.

In order to get rid of residua and to measure the transport properties, the as-synthesized polycrystalline sample was pressed into pellets and annealed at the temperatures from 120 °C to 200 °C and under the high pressure from 2 GPa to 5 GPa for 5 hrs. X-ray diffraction (XRD) was performed by using X-ray diffractometer (SmartLab-9, Rikagu Corp.) with Cu Kα radiation and a fixed graphite monochromator. The XRD patterns were collected in the 2-theta range of 5°–80° with a scanning rate of 0.1°/min at room temperature. The lattice parameters and the crystal structure were refined by using the Rietveld method in the programs

GSAS package, applying Thompson-Cox-Hastings functions with asymmetry corrections as reflection profiles. The magnetization measurements were made by SQUID MPMS (Quantum Design). The magnetizations under pressure were measured by incorporating a copper–beryllium pressure cell (EasyLab) into SQUID MPMS. The sample was firstly placed in a teflon cell (EasyLab) with Daphne Oil 7373 as the pressure media. Then, the teflon cell was set in the copper–beryllium pressure cell for magnetization measurement. The contribution of background magnetization against the sample at 2 K was less than 5%. The resistivity measurements were performed using a Quantum Design physical properties measurement system (PPMS-9).

**Notes:** For this work, we planned to synthesize MgOFeSe, so that we refined the XRD pattern by using MgOFeSe structure model and got a quite good $R_{wp}$. Products in high-temperature decomposition seem to include MgO and FeSe. Therefore, we believed that MgOFeSe was responsible for superconductivity at 42 K in the first version. But in the further research with no Mg element adding as the starting material for the synthesis, we still observed superconductivity at almost the same temperature, and nearly the same XRD pattern was collected. We refined the XRD pattern adopting $LiFeO_2Fe_2Se_2$ structure model, lower $R_{wp}$ (9.37%) was obtained. At the meanwhile, high-temperature

decomposition gave rise to α-LiFeO$_2$, whose XRD pattern is very similar to MgO, and Fe selenides. Further, the composition of superconducting phase was determined to be LiFeO$_2$Fe$_2$Se$_2$ by ICP-AES. We are sorry for this confusion.

**Acknowledgments**

The authors are grateful for the discussions with Z. Sun and T. Wu. This work is supported by the National Natural Science Foundation of China (Grants No. 11190021, 51125006, 91122034), and the "Strategic Priority Research Program (B)" of the Chinese Academy of Sciences (Grant No. XDB04040100), and the National Basic Research Program of China (973 Program, Grant No. 2011CBA00101).

**Author contributions**

X.F.L. and N.Z.W. performed sample synthesis and all experiments with assistance from X.G.L. and B.L.. G.H.Z. did the Rietveld refinement, and Z.M.M. and X.F.L. proposed the synthetic route. X.F.L. and X.G.L. analyzed the data. X.H.C. and X.G.L. wrote the paper. F.Q.H. and X.H.C. conceived and coordinated the project. X.H.C. is responsible for the infrastructure and project direction. All authors discussed the results and commented on the manuscript.

**Additional Information**

The authors declare no competing financial interests. Correspondence and requests for materials should be addressed to huangfq@mail.sic.ac.cn and chenxh@ustc.edu.cn.

**Table 1.** The crystallographic parameters for the Rietveld refinements of ($Li_{0.5}Fe_{0.5}$)OFeSe from XRD data [a].

| Atom | x | y | z | g |
| --- | --- | --- | --- | --- |
| Li | 0.25 | 0.75 | 0 | 0.5 |
| Fe1 | 0.25 | 0.75 | 0 | 0.5 |
| Fe2 | 0.75 | 0.25 | 0.5 | 1 |
| O | 0.25 | 0.25 | -0.0764(1) | 1 |
| Se | 0.25 | 0.25 | 0.3355(1) | 1 |

Bond length (Å)

| | |
| --- | --- |
| Fe2-Se | 2.4347(1)×4 |
| Fe2-Fe | 2.6818(1)×4 |

Bond angle (deg)

| | |
| --- | --- |
| $Se^{Top}$-Fe2-$Se^{Top}$ | 102.31(1)×2 |
| $Se^{Top}$-Fe2-$Se^{Bottom}$ | 113.16(1)×4 |

[a] Space group: *P4/nmm* (No. 129); $a = b$ = 3.7926(1) Å, $c$ = 9.2845(1) Å, $V$ = 133.54(1) Å$^3$, $R_{wp}$ = 0.0937, $R$ = 0.0656, $\chi^2$=1.43. $g$ is the occupation factor.

**Figure Legends:**

**Figure 1| Structural view and Rietveld refinement patterns for $(Li_{0.5}Fe_{0.5})OFeSe$.** **(a):** A schematic view of the structure of $(Li_{0.5}Fe_{0.5})OFeSe$. The layered structure consists of alterative stacking anti-PbO-type FeSe layers and $(Li_{0.5}Fe_{0.5})O$ layers. **(b):** XRD patterns together with Rietveld refinement results. The observed diffractions intensities are represented by yellow-green open circles, and the fitting pattern by the red solid line. The blue curve at the bottom represents the difference between observed and fitting intensities. Short black vertical bars below the XRD patterns indicate the positions of allowed Bragg reflections.

**Figure 2| Magnetic susceptibility χ of $(Li_{0.5}Fe_{0.5})OFeSe$.**
**(a)**: The magnetic susceptibility of the as-synthesized sample from hydrothermal method. **(b)**: The magnetic susceptibility of the sample after annealing under high pressure of 5 Gpa and at 150 °C for 5 hours. Filled symbols: zero-field cooling (ZFC); open symbols: field cooling (FC). These data were collected under the magnetic field of 10 Oe. The inset in (a) is the zoomed view around the superconducting transition of the as-synthesized sample, which is around 40 K. The bottom inset of (b) is the zoomed view for ZFC data around the superconducting transition of the annealed sample, which indicates $T_c = 43$ K. The top inset of (b) is the

*M-H* loop taken at 5 K for the annealed sample, where a linear H dependence of *M* with a negative slope can be recognized below about 300 Oe in low-field range (as the magenta arrow points).

**Figure 3| Pressure dependence of the superconductivity in $(Li_{0.5}Fe_{0.5})OFeSe$.** **(a)**: the magnetizations (*M*) taken under various external pressures ranging from 0 – 1 GPa and H = 10 Oe in ZFC mode. The measured sample was annealed under high pressure of 5 GPa and at 150 °C for 5 hours. The inset shows the zoomed view around the superconducting transition. The arrow shows the beginning of the drop of *M* at 1 GPa, which represents the way to define $T_c$ shown in the bottom panel. **(b)**: Pressure dependence of the superconducting transition temperature, $T_c$.

**Figure 4| $T_c$ as function of chalcogen (Se) height for the FeSe-derived superconductors.** The data of FeSe under high pressure are indicated by blue filled squares (from Ref. 31). Red filled circles indicate the data of the intercalated FeSe superconductors. The data of $K_xFe_{2-y}Se_2$ and $Li_{0.6}(NH_2)_{0.2}(NH_3)_{0.8}Fe_2Se_2$ come from Ref. 7 and Ref. 30, respectively. The data of $LiFeO_2Fe_2Se_2$ is from the present work.

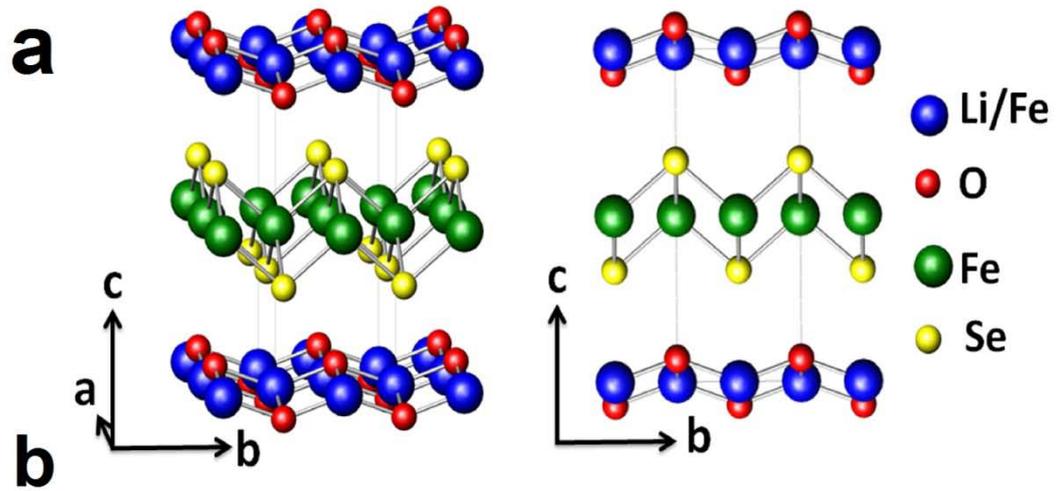
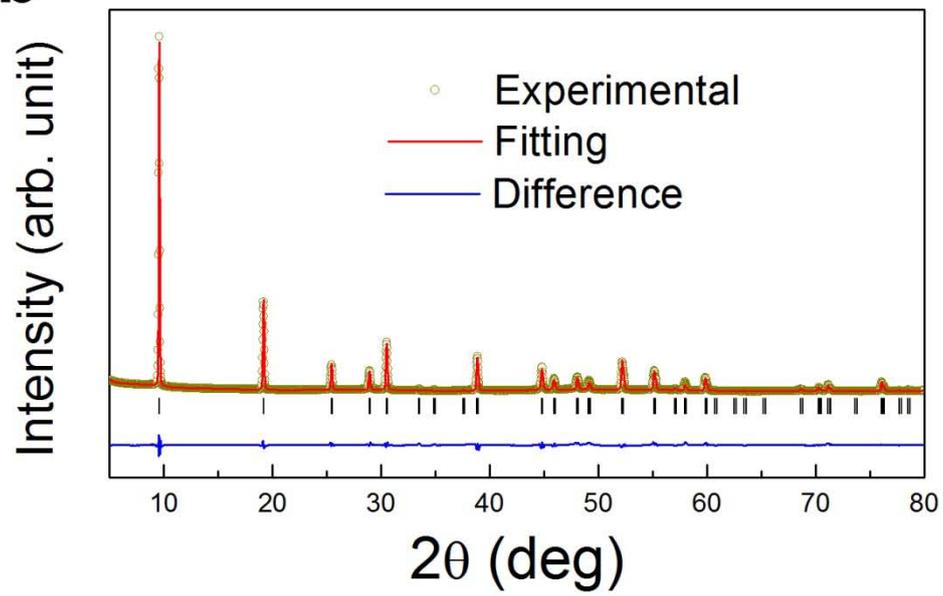

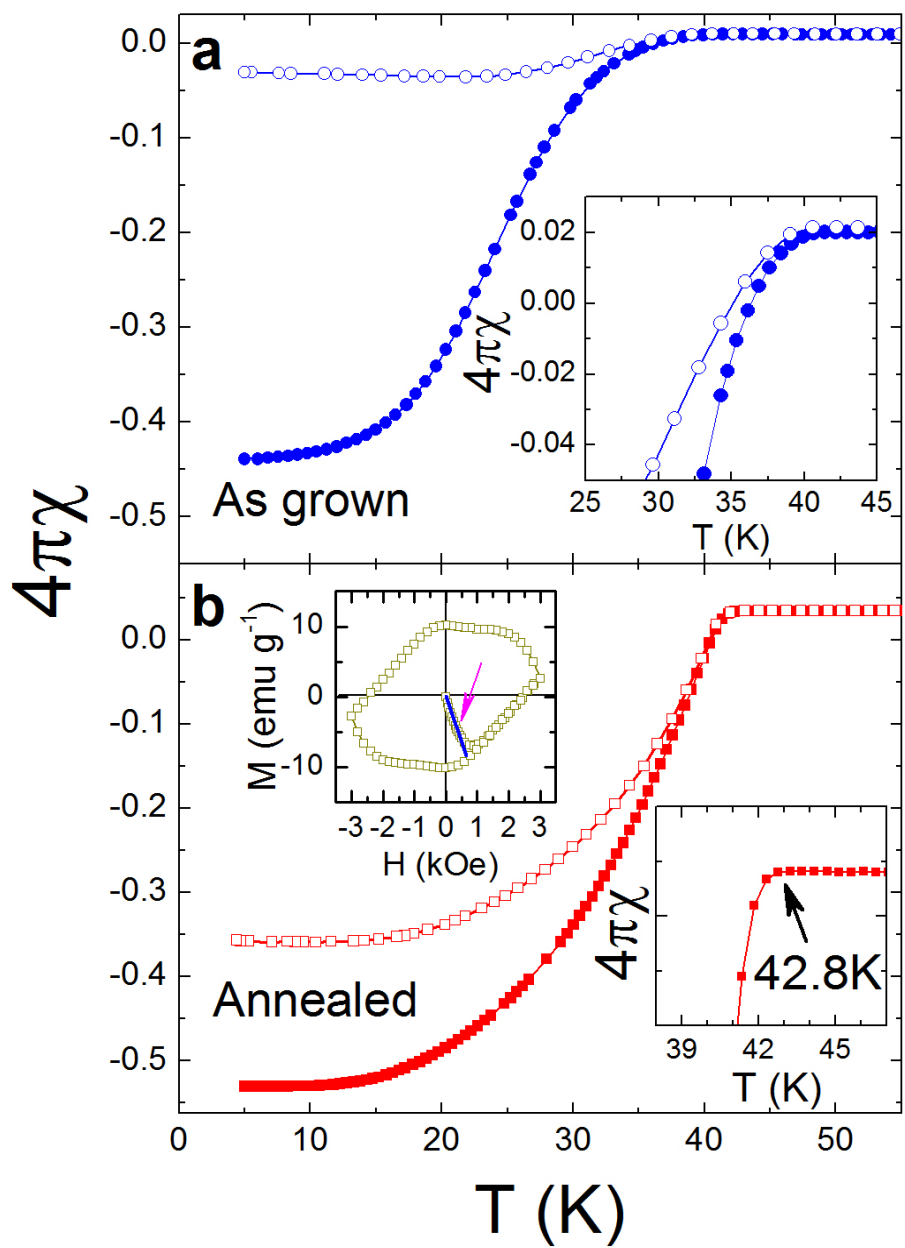

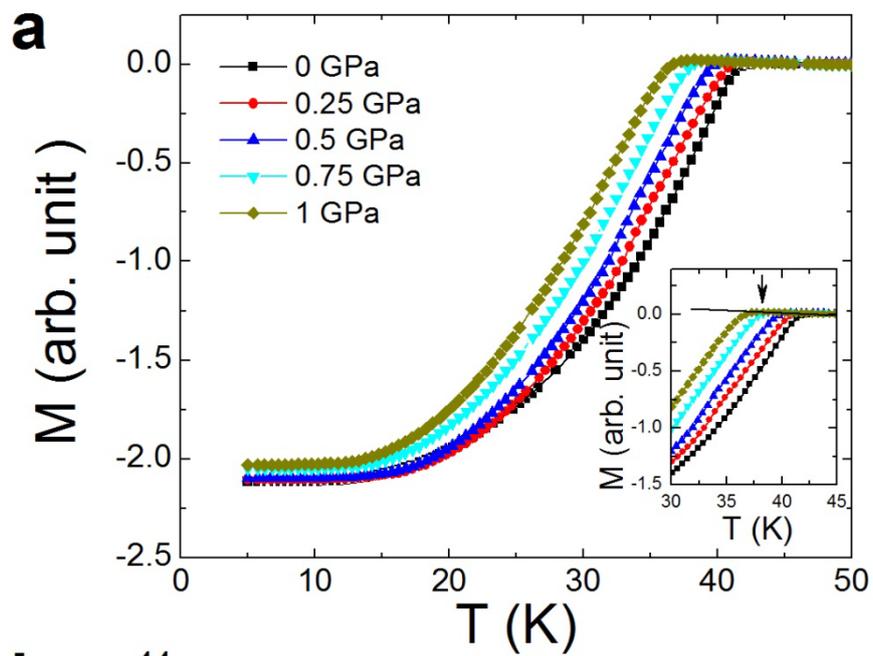

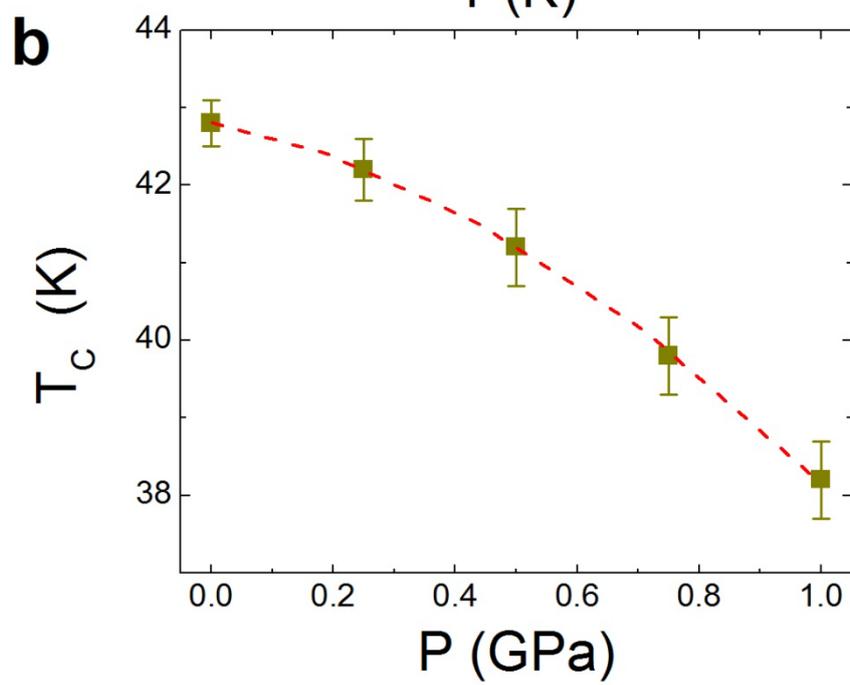

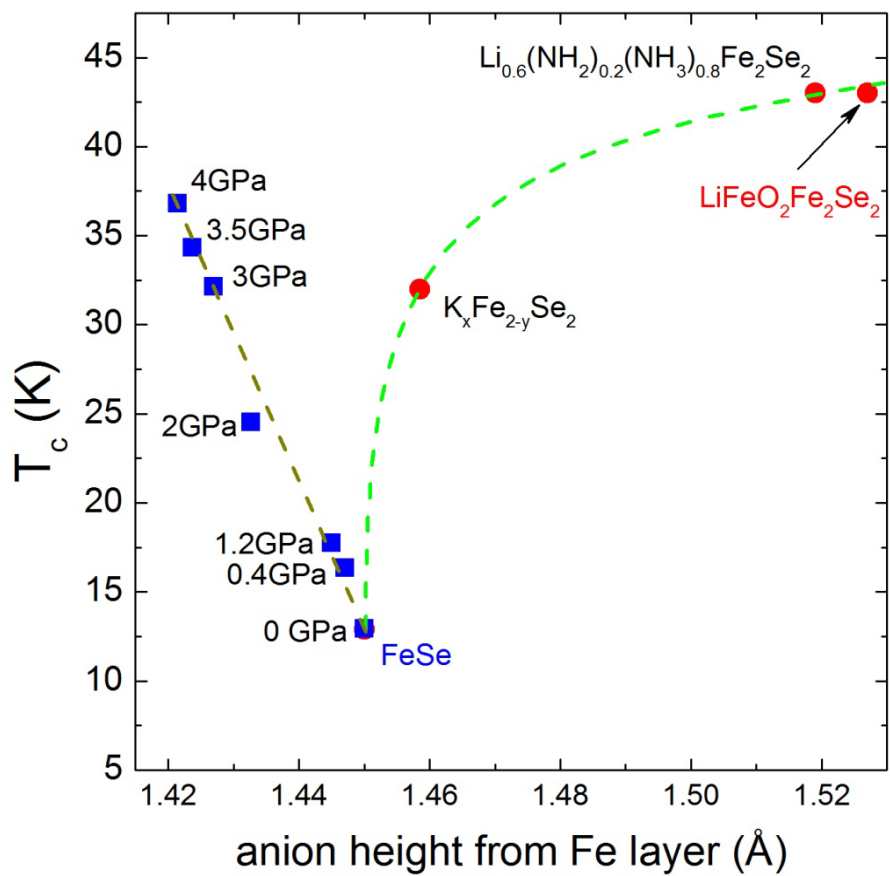